\documentclass[]{spie}
\usepackage[]{graphicx}
\title{Rotation Dynamics of a Galaxy\\ with a Double Mass Distribution}
\author{Louis Marmet}
\authorinfo{Author information: e-mail to ``myfirstname''@marmet.org .\\Copyright December 28$^{th}$, 2005. Version dated November 19$^{th}$, 2007.\\}

\begin{document}
\maketitle

\begin{abstract}
The rotation dynamics of spiral galaxies is modeled using a sum of two mass distributions: a spherical bulge and a thin disk. The density functions representing these mass distributions are calculated from the total angular momentum of the galaxy and known rotation curves. When both bulge and disk density functions are assumed to be smooth and are given a value of zero beyond the edge of the galaxy, a unique solution is obtained. Moreover, the calculated density functions show that constant rotation curves are obtained from a nearly exponential luminosity profile with a dark matter distribution which follows that of the light emitting matter, without the need for modified Newtonian dynamics.
 Experimental detection of molecular hydrogen in spiral galaxies confirms the presence of a baryonic massive component consistent with the results obtained with this model. It is proposed that dark matter does not have to be exotic non-baryonic matter. Instead, it can be made of molecular hydrogen and condensed matter.
\end{abstract}

\keywords{galaxy rotation curve, spiral galaxy, bulge, disk, dark matter}

\section{Introduction} \label{sec:introduction}
Spiral galaxies have rotation curves which seem defy any attempt at explanation by simple celestial mechanics. In a Keplerian system where most of the mass is localized at the center, the orbital velocity decreases as the inverse of the square root of the distance from the center. However, the rotation curves of spiral galaxies are typically constant over the entire radius of the galaxy, except near the center where an approximately linear dependence on the radius is observed. The nearly constant rotation speed has been observed experimentally in most spiral galaxies\cite{Rubin1993}. A simple argument shows that the constant velocity curves imply that the cumulative mass of the galaxy must rise linearly with radius. The argument is as follows: The equation of motion of a planet at a radius $r$ in a circular orbit is $GM/r^2 = v^2/r$. From Newtonian mechanics, ``the force on a mass at radius $r$ from the center of a symmetrical mass distribution is proportional to the mass interior to that $r$''.\cite{Rubin2006} Also, the gravitational field outside a spherical shell is as if the mass of the shell were concentrated at its center. Therefore, the equation of motion can be simplified to $GM(r) = v^2r$, where $M(r)$ is the mass interior to the radius $r$. Since the mass outside the radius $r$ does not have any gravitational effect, and since $v$ is independent of $r$ as per experimental measurements, one must have $M(r) \propto r$. This conclusion does not agree with the visible mass distributions of galaxies, which instead show a mass density which drops exponentially as a function of the radius.

To explain the constant rotation curves, more complex mass distributions have been examined. Analytical functions\cite{Binney1987} were chosen to imitate the intensity profiles of galaxies, a choice based on the assumption that the mass-to-luminosity ratio is independent of the radius within a given galaxy. Functions such as the de Vaucouleurs profile and the S\'ersic profile were used for the mass density of the bulge, while an exponential profile was used for the mass density of the disk. The calculation of the rotation curve was obtained from an integral over the radius of the gravitational contribution of every mass element taken from zero to infinity. However, none of these profiles reproduces a flat rotation curve, a conclusion that suggests that the mass distributions in galaxies does not follow the intensity profile. If an additional mass distribution is added to the density function obtained from the the intensity profiles,\cite{vanAlbada1985, Rubin1993} a constant rotation curve can be obtained. Such hypothetical distribution constitutes the third component of a galaxy, the dark matter halo, which is detectable only from its gravitational effects on the galaxy dynamics. Because the halo emits a small amount of light, it is thought to be mostly made of dark matter.

\section{Other Mass Distributions}
Apart from gravitational effects, dark matter has never been detected. A question arises whether other mass distributions, not considered in the simple argument above or approximated by analytical functions, could produce a flat rotation curve. Such distributions would also have to be compatible with the known mass of galaxies. The first step in answering this question is to show that the simple argument of a Keplerian system cannot be applied to a spiral galaxy.

It is established that a mass inside a spherical shell experiences no net gravitational force from that shell. This is a special case of the general properties of a homeoid: a shell of uniform density between similar, concentric ellipsoids. In general, a mass located inside a homeoid experiences no net gravitational force from that homeoid. The gravitational field outside a spherical shell is as if the mass of the shell were concentrated at its center. In general, the gravitational force exerted on the mass located at radius $r$ will be proportional to the mass of all the similar homeoids interior to $r$. However, the gravitational force outside a \emph{non-spherical} homeoid is not described by an inverse square law dependence on the radius. Spiral galaxies clearly require the sum of non-spherical homeoids in order to reproduce their disk-like geometry. The equation of motion must be replaced by the more general $Gf(M(r), r) = v^2/r$, where $f$ is a function involving elliptic integrals. The expected conclusion - that the cumulative mass of an elliptical galaxy rises linearly with radius - is incorrect.\cite{Marmet2007}

To expand on the set of functions that were treated analytically, numerical methods are used to study galaxy dynamics. Mass distributions can be numerically integrated to find the corresponding rotation curve, and reverse algorithms have been developed to obtain the mass distributions from arbitrary rotation curves. Some numerical models use a mass distribution entirely located on the galactic disk\cite{Mera1997} or on a disk with a variable thickness\cite{Nicholson2000}. One advantage of the numerical method is that a given rotation curve can be inverted to find a unique mass distribution.

This paper proposes to use two mass distributions to simulate the bulge and the disk of spiral galaxies. These two components of spiral galaxies produce fundamentally different gravitational fields. The two distributions are clearly apparent on images of galaxies taken in the visible and other parts of the spectrum. The bulge is a near spherical distribution of stars (Population II system) believed to be almost as old as the galaxy. By contrast, the disk stars (Population I system) have formed at a steady state rate during the evolution of the galaxy. More heavy elements are found in the disk, which often shows more absorption of light when seen edge-on. A spherical bulge produces a centrally symmetric gravitational force proportional to $M(r)/r^2$. Spherical shells of matter outside $r$ do not contribute to the gravitational force. A disk produces a gravitational force described by less intuitive elliptic integrals. The force on a point located at the cylindrical coordinate $R$ is a function of the mass inside $R$, but unlike the case of spherical shells, circular rings of matter outside $R$ contribute to the gravitational force in a direction pointing away from the center of the system. The dynamics of a galaxy with bulge and disk components can be calculated by using a disk with a varying thickness\cite{Nicholson2000} given by the function $h(R)$. For a given function $h(R)$ and a rotation curve $v(R)$, a unique mass density $\rho(R)$ is obtained. However, the size of the bulge relative to the size of the disk must be known to determine the function $h(R)$.

The problem can be addressed differently with a mass density given by the sum $\rho(\vec r) = \rho_B(\vec r) + \rho_D(\vec r)$, where $\rho_B(\vec r)$ is the mass distribution of a spherical bulge, and $\rho_D(\vec r )$ is the mass distribution of a thin disk. Without \emph{a priori} knowledge of the relative sizes of the bulge and disk, this double mass distribution gives the densities of the two galaxy components needed to produce a given velocity curve $v(R)$.\cite{Gallo2006}

\section{Theory} \label{sec:theory}
The galaxy is assumed to have a rotation axis $\vec\Omega$ and a galactic plane perpendicular to the rotation axis. The center of the galaxy is at the intersection of the rotation axis and the galactic plane. The mass density is $\rho(\vec r) = \rho_B(\vec r) + \rho_D(\vec r)$, where $\vec r$ describes the position from the center of the galaxy. The density is assumed to be independent of the azimuthal angle and can therefore be represented by the two-dimensional function $\rho(R, z)$, where $z$ is the height above the galactic plane. This restriction implies that some details such as the arms of a spiral galaxy do not appear in the model. A spherical symmetry is assumed for the bulge, with the density being $\rho_B(r)$. The disk is considered thin $h << R_{max}$, with a surface mass density given by $h\rho_D(R)$, where $h$ is a constant representing an equivalent thickness. The density distribution extends to a radius $R_{max}$, beyond which it falls to zero $\rho(r > R_{max}) = 0$.

The system is assumed to be in a steady state of motion with $\rho(R, z)$ independent of time. The velocity is described by a function $\vec v(R)$ parallel to $\vec\Omega\times\vec r$. The matter thus describes circular paths of radius $R$ around the galaxy's rotation axis. For simplicity, the velocity is chosen to be independent of $z$. The contribution of the centrifugal force which would likely produce a flattened ellipsoid is neglected for the bulge.

The angular momentum of the mass at $\vec r$ is given by:
\begin{equation} 
\vec L(\vec r) = \rho(\vec r) \vec r \times \vec v(\vec r) 
\label{eq:ang_momentum}
\end{equation}
and the total angular momentum of the system $\vec L$ is obtained by integrating $\vec L(\vec r)$ over the volume of the entire mass distribution. It is also useful to define the intrinsic angular momentum
\begin{equation} 
\Lambda = L / (M_{galaxy} R_{max}).
\label{eq:intrinsic_ang_momentum}
\end{equation}
which gives a representation of the amount of mass located in the regions where the rotation velocity is large.

{\bf Forces}

Two types of forces are considered: gravitational and centrifugal. The gravity from a mass located at $\vec r_0$ produces an acceleration $\vec a(\vec r, \vec r_0)$ at point $\vec r$ given by Newton's law of universal gravitation:
\begin{equation} 
\vec a(\vec r, \vec r_0) = -G\rho(\vec r_0) (\vec r - \vec r_0) / |\vec r - \vec r_0 |^3 .
\label{eq:newtons_law}
\end{equation}
where $G$ is the gravitational constant. A modified Newtonian dynamics\cite{Milgrom1983} (MOND) is not used here. The acceleration at a point $\vec r$ produced by a spherical shell of uniform density $\rho_B(r_s)$ is given by:
\begin{equation} 
\vec a(\vec r, r_s) = -4\pi G \rho_B(r_s) r_s^2 \vec r/r^3
\label{eq:acceleration}
\end{equation}
for $r_s\leq |\vec r|$, otherwise $\vec a(\vec r, r_s) = 0$. The total acceleration from the bulge is obtained by integrating $\vec a(\vec r, r_s)$ over all radius elements $dr_s$. In the case of a disk, Eq. (\ref{eq:newtons_law}) is integrated numerically (see Appendix A).

The centrifugal acceleration at radius $R$ is:
\begin{equation} 
a(R) = \omega^2(R) R = v^2(R)/R
\label{eq:centrifugal_acc}
\end{equation}
The rotation curve is obtained by balancing the gravitational acceleration with the centrifugal acceleration in the plane of the disk.

All other types of forces and effects are neglected. Other models describing galaxy dynamics in the early stages of formation include electromagnetic forces as a result of the high plasma density\cite{Peratt1986a, Peratt1986b, Peratt1990}. Both thermal and radiation pressures are neglected. Relativistic effects are also neglected, the gravitational interaction is assumed to be instantaneous across the size of the galaxy. Furthermore, no stability analysis is done on this model. These approximations are discussed below.

\section{Numerical inversion of the rotation curve} \label{sec:numericalcalculation}

A numerical calculation is used to evaluate the required density of matter to produce the target velocity profile $v_0(R)$ (Fig. \ref{fig:velocityprofiles}) with an intrinsic angular momentum $\Lambda_0$. The density profiles of the bulge and the disk are numerically represented by two functions $\rho_B(r_i)$ and $\rho_D(r_i)$ for $N$ equally spaced points along the radius. An initial mass distribution is given to the program which then follows these steps:

\begin{figure} 
\begin{center}Rotation Curves\\ \end{center}
\centerline{\includegraphics[width=6.25in]{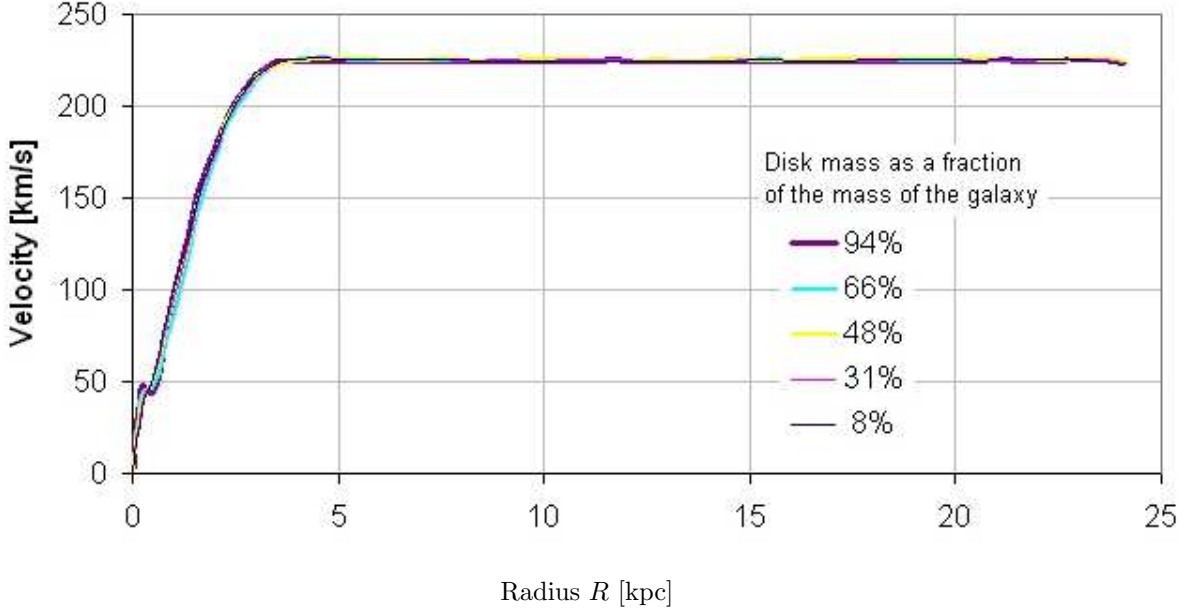}}
\begin{center}Radius $R$ [kpc]\\ \end{center}
\caption{Velocity profiles of hypothetical galaxies with different angular momentum. The calculated mass distributions produce rotation curves that are very close to the target velocity function.}
\label{fig:velocityprofiles}
\end{figure}

\begin{enumerate} 
    \item \label{it:randomvar} A small random variation of size $\pm dM$ is added to the mass density profile. Negative mass values are not allowed.
    \item The gravitational acceleration is calculated in the plane of the disk at $\phi = 0$ for all values of the radius $r_i = (i-1) \Delta r$ ($i = 1$ to $N$). The acceleration produced by the bulge is obtained from Eq. \ref{eq:acceleration} by summing the contributions of all $N$ shells of radius $r < r_i$. The accelerations from each rings are obtained by first integrating the mass elements on each ring (at coordinates $r$, $\phi$), and then summing the acceleration from every rings. The acceleration from the mass element at $r = r_i$, $\phi = 0$, is considered to be zero by symmetry since by choice, $r_i$ is located at the center of the mass element. This simple solution avoids all the problems associated with an infinitely thin disk and divergent integrals.
    \item From the total acceleration, the required velocity is found with Eq. \ref{eq:centrifugal_acc}.
    \item The deviation is calculated between the calculated and the target velocity curves.
    \item \label{it:improvedfit} If the random variation has improved the fit, the new density profile is kept, otherwise the program reverts back to the old profile.
    \item \label{it:smooth} When about one hundred iterations of steps \ref{it:randomvar} through \ref{it:improvedfit} have been done, the density profiles $\rho_B(r)$ and $\rho_D(r)$ are smoothed with a low pass filter $\rho '(r_i) = (\rho(r_{i-1}) + S_s\rho(r_i) + \rho(r_{i+1}) )/(S_s+2)$, with $i = 2$ to $N-1$, and $S_s$ being the smoothing strength. The central values $\rho_B(0)$ and $\rho_D(0)$ are not affected by the smoothing. For the values at the largest radius $\rho(r_N) = (\rho(r_{N-1}) + S_s\rho(r_N) )/(S_s+2)$ is used. This smoothing simulates a diffusion process between consecutive rings and shells. For systems with low intrinsic angular momentum, the value $S_s = 3$ is used. However, systems with high intrinsic angular momentum have a faster varying disk density distribution, and a value as large as $S_s = 25$ is required.
    \item The intrinsic angular momentum of the system is calculated and compared to the target angular momentum $\Lambda_0$. If the system's intrinsic angular momentum is too small, some mass $a_l\rho_B(r)$ is transferred from the bulge to the disk, otherwise some mass $a_l\rho_D(r)$ from the disk is transferred to the bulge. The fraction $a_l$ is of the order of a few percent, and smaller if the intrinsic angular momentum is already near the target $\Lambda_0$. Since the mass transfer changes the gravitational field of the system, steps \ref{it:randomvar} through \ref{it:smooth} have to be repeated again.
\end{enumerate}

After several iterations, the magnitudes of $dM$ and $a_l$ are slowly reduced, and $S_s$ is slightly increased until a good fit is obtained for the rotational curve and the intrinsic angular momentum. The source code is available\cite{Marmet2006}.

\begin{figure} 
\begin{center}Mass Densities for the Bulge and Disk of a S0 galaxy\\ \end{center}
\centerline{\includegraphics[width=6.25in]{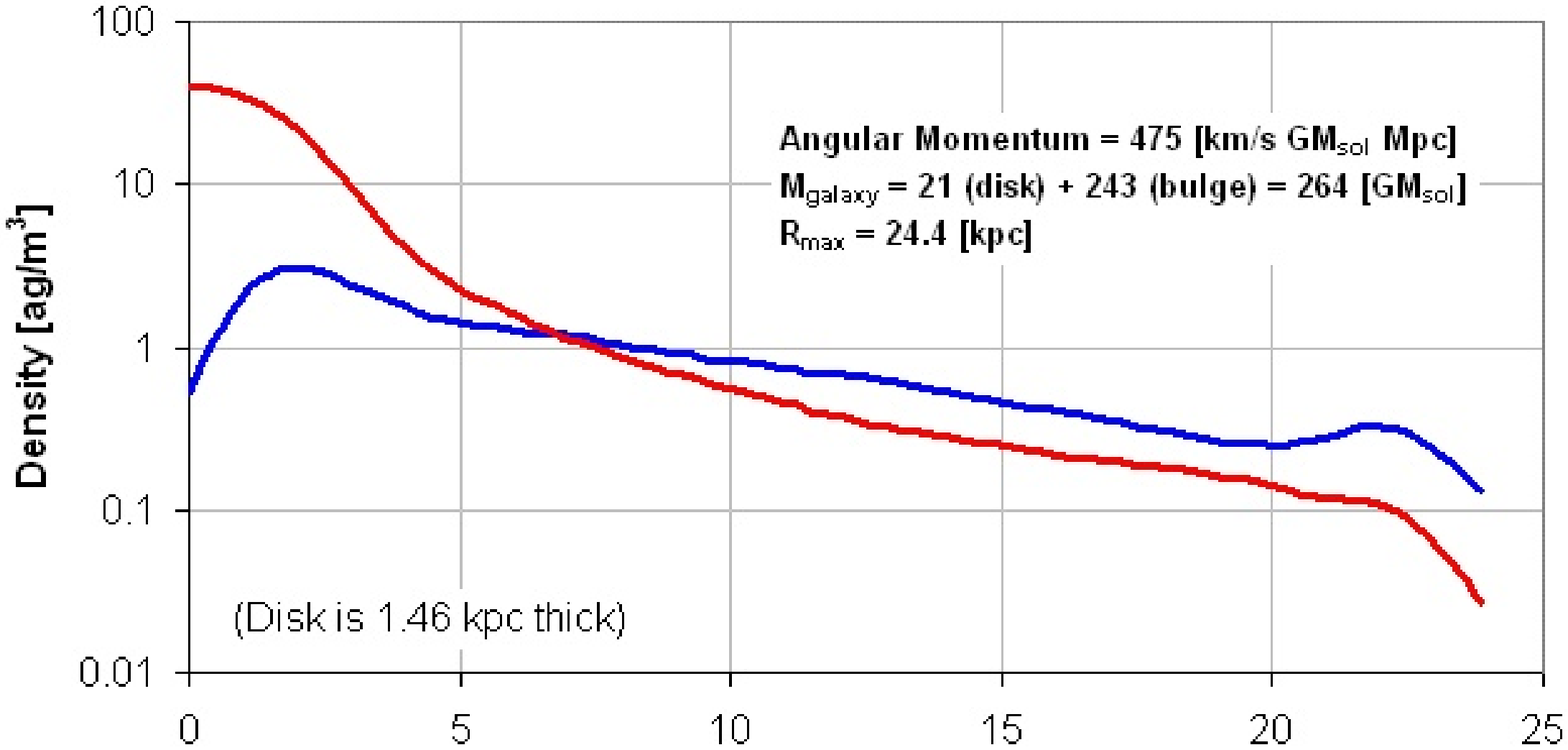}}
\begin{center}Radius $R$ [kpc]\\ \end{center}
\caption{Density profile of a hypothetical galaxy represented by the sum of a spherical (red) and a thin disk (blue) mass density distribution. The intrinsic angular momentum is near the smallest possible value $\Lambda_{min} = 73.8km/s\ M_{galaxy}R_{max}$ for the given rotation curve. $92\%$ of the mass of the galaxy is in the bulge.}
\label{fig:density7385}
\end{figure}

\begin{figure} 
\begin{center}Mass Densities for the Bulge and Disk of a Sd galaxy\\ \end{center}
\centerline{\includegraphics[width=6.25in]{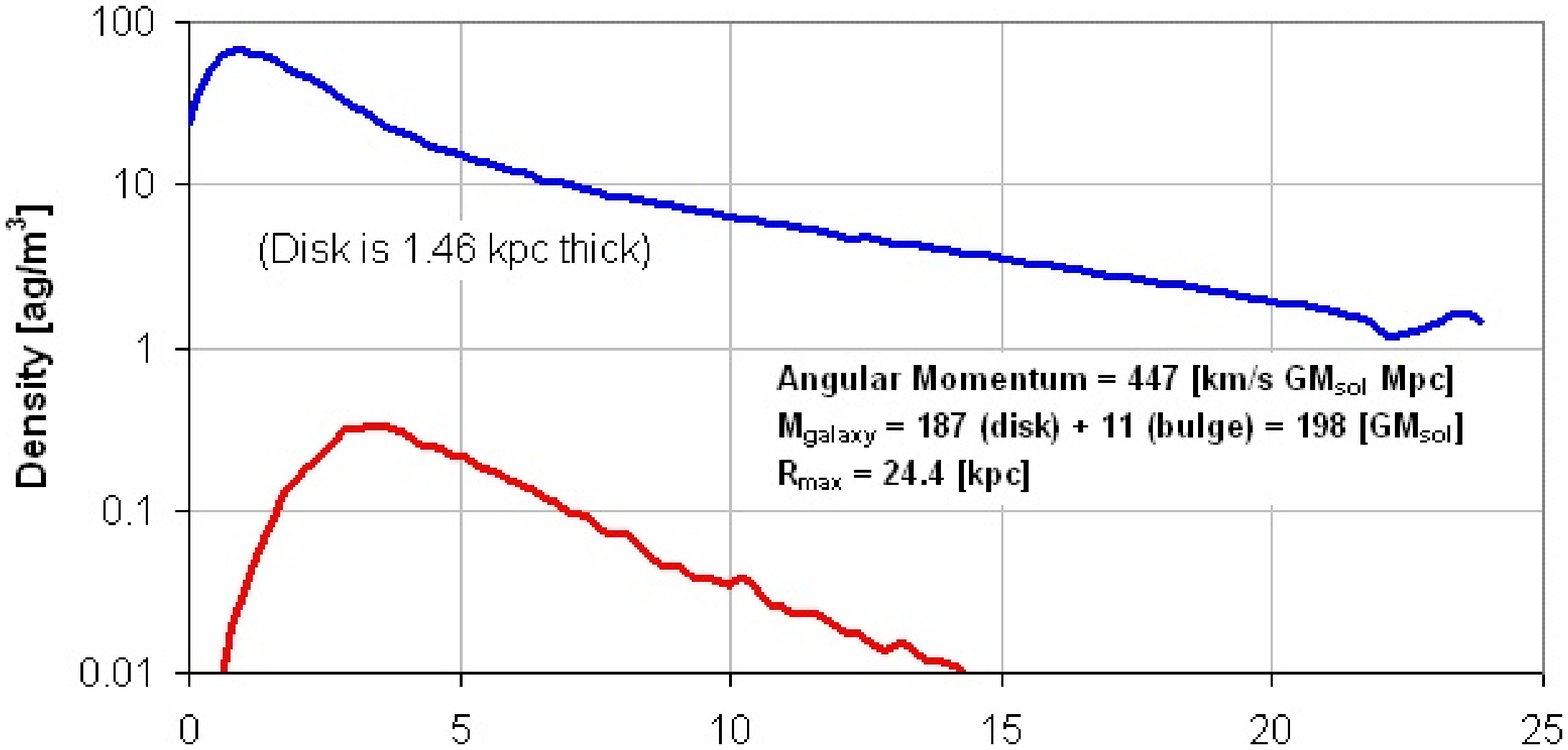}}
\begin{center}Radius $R$ [kpc]\end{center}
\caption{Density profile of a hypothetical galaxy represented by the sum of a spherical (red) and a thin disk (blue) mass density distribution. The angular momentum is near the highest possible value $\Lambda_{max} = 92.7km/s\ M_{galaxy} R_{max}$ for the given rotation curve. $94\%$ of the mass of the galaxy is in the disk.}
\label{fig:density9264}
\end{figure}

As an example, the following boundary conditions are used. A typical size of $R_{max} = 24.4kpc$ is given to the galaxy. The disk thickness is chosen at $h = 1.6kpc$. The initial velocity curve has the profile:
$$v(R) = 225km/s \times (1-(3kpc-R)^2/(3kpc)^2), for R < 3kpc,$$
$$v(R) = 225km/s,\ \ \ \ \ \ \ \ \ \ \ \ \ \ \ \ \ \ \ \ \ \ \ \  for R > 3kpc.$$
This velocity curve has a linear increase for small $R$ and smoothly converges to a constant $225km/s$ for larger radii. Given this rotation curve, the intrinsic angular momentum can take any value between $\Lambda_{max} = 73.8km/s\ M_{galaxy}R_{max}$ (all bulge) and $\Lambda_{min} = 92.7km/s\ M_{galaxy}R_{max}$ (all disk). The angular momentum varies between $L_{min} = 442km/s\ GM_{sol}Mpc$ ($66\% $ mass in disk) and $L_{max} = 475km/s\ GM_{sol}Mpc$ (all mass in bulge). The results for these extreme cases of the intrinsic angular momentum are shown in Figs. \ref{fig:density7385} and \ref{fig:density9264}.


For $\Lambda_{min}$, the mass of the galaxy is $2.64\times 10^{11} M_{sol}$ while for $\Lambda_{max}$, the mass is the smallest at $1.98\times 10^{11} M_{sol}$. This reflects the larger mass of the bulge required to produce the same gravitational acceleration as that produced by the disk.

The rotation curves obtained with different values for the angular momentum (Fig. \ref{fig:velocityprofiles}) show excellent fits as seen by the good overlap of the five curves. Small deviations are observed near $R = 0.5kpc$ for the high-$\Lambda$ ``disk'' galaxies, indicating that a linear increase of the velocity is difficult to model with a disk only. As expected, the gravitational acceleration increases (in absolute value) with radius up to $R = 2.5kpc$, then decreases as $1/R$, cancelling the centrifugal acceleration determined by $v(R)$.

\begin{figure} 
\begin{center}Column Density (Bulge and Disk)\\
\centerline{\includegraphics[width=6.25in]{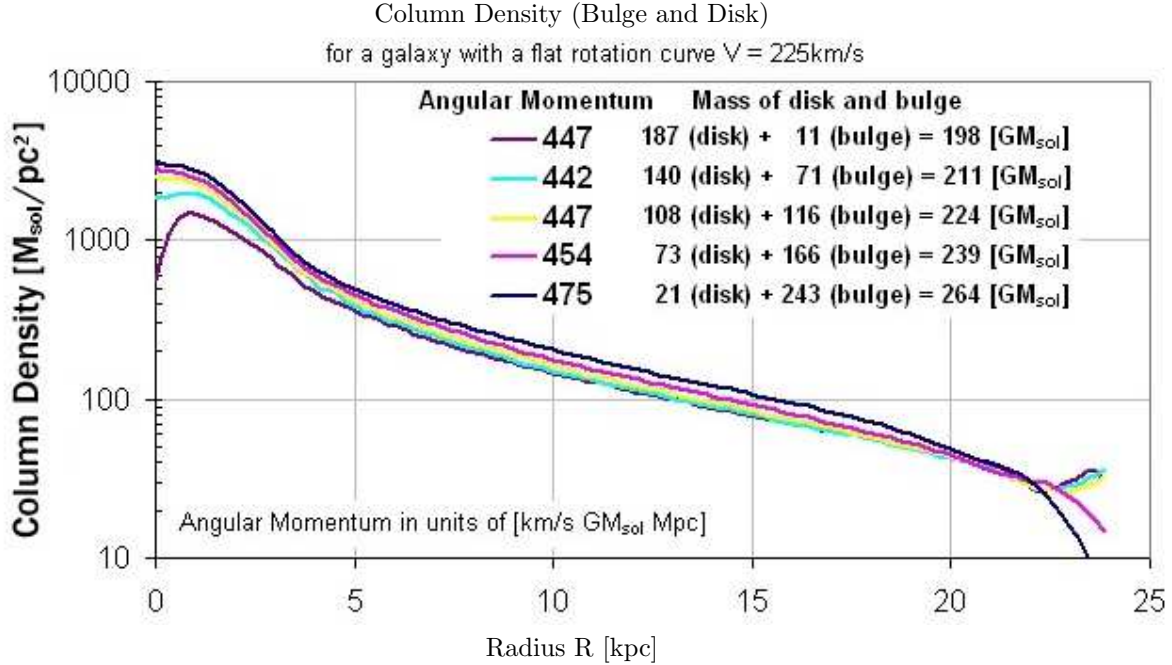}}
Radius R [kpc]\\
\end{center}
\caption{Column density of the hypothetical galaxy for different values of the angular momentum. The logarithm of the density is plotted as a function of the radial position in the galaxy.}
\label{fig:columndensity}
\end{figure}

A $24 kpc$ radius galaxy with a $225 km/s$ rotation velocity will weigh about $220 GM_{sol}$. The column density is calculated by integrating the mass density through the thickness of the galaxy. The result is shown in Fig. \ref{fig:columndensity}. The column density does not vary by more than a factor two for all the possible values of the angular momentum. A mass density of $1500$ to $3000 M_{sol}/pc^2$ is in agreement with the measurements of Valentijn et al.\cite{Valentijn1999}.

\section{Mathematical considerations} \label{sec:discussmath}

Using the numerical calculations, it is possible to obtain a flat rotation curve from either a spherical or a disk distribution of matter. The numerical solutions are however physically meaningful if the mass distribution is limited to a finite radius, otherwise the total mass of the galaxy does not tend to converge. Without any additional conditions, there still exists an infinite number of solutions which will give the target rotation curve, each involving a different bulge-to-disk mass ratio. In the calculations above, smooth functions of the radius are used and the angular momentum of the system is defined as an initial restriction. In this case, the problem has a unique solution.

{\bf Inversion of the rotation curve}

The solutions obtained above correspond to the ``inversion'' of the rotation curve. The inversion process assumes a specific type of mass distribution. If a solution represented by a spherical distribution is assumed, the inversion of the rotation curve produces a unique solution for the mass distribution. However, not all solutions are physically significant since negative density values might be obtained. More specifically, the rotation curve $v(R)$ can be ``inverted'' analytically to obtain the spherical mass distribution\cite{Mizony2003} $\rho(R) = 1/(4\pi R^2)\ d/dR\ (R\ v^2(R))$. As long as the derivative of $v(R)$ with respect to $R$ is not ``too negative'' such that $dv(R)/dR > -v(R)/2R$, there is a positive density solution for the spherical distribution.

If a disk distribution is assumed, the rotation curve $v(R)$ is not easy to invert analytically. Numerical techniques become almost essential to treat this case. However, again, some rotation curves yield an unphysical negative mass density. For example, a linear increase of the velocity at small radii causes the density to become negative. This happens for the $94\%$ disk galaxy where the rotation curve cannot be made linear near the center, as is visible in Figure \ref{fig:velocityprofiles}. (The current algorithm does not allow negative values of the density. The best numerical solution gives values of the density that are near zero.) However, the exact behaviour of the velocity curve near the center is not too critical, and it is also very likely masked by the light emitted by the bulge. The strange behaviour near the center (Fig. \ref{fig:columndensity}) is only a result of the incorrect choice of the rotation curve near the center.

If a double mass distribution is assumed, the numerical description is represented by $2N$ unknowns. The mass can be distributed either on the disk or the spherical bulge (with all possible combinations of ratios at different radius) while obtaining the same target rotation curve. In order to uniquely specify the solution, additional independent equations are needed. As with the other two cases, there are $N$ independent equations generated by balancing the centrifugal force with the gravitational force at every radii $r_i$. Another $N$ independent equations appear from the one additional constraint on the angular momentum and the smoothing condition which produces $N-1$ equations. The physical reason behind the smoothing is that matter diffuses from one ring or shell to the next one. The diffusion introduces a new physical mechanism which is associated with the $N-1$ new independent mathematical equations. There are therefore enough independent equations to completely specify the problem.

\section{Physical considerations} \label{sec:discussphys}

The double mass distribution gives new results about the physics involved in the dynamics of a galaxy. These are discussed here.

{\bf Finite extent of the mass distribution}

The mass distribution of a galaxy is usually inferred from the velocity measurements\cite{Burstein1982}. The usual conclusion for a spherical distribution is that if $v(r)$ is constant, $M(r)$ must increase linearly with $r$ since $v(r) = \sqrt{GM(r)/r}$. This is the case if the density of the spherical bulge is $\rho_B(r) \propto 1/r^2$. However, if no limit is assumed on this density distribution, the mass of the galaxy becomes infinite.

Binney and Tremaine\cite{Binney1987} assume that the density distribution follows an exponential law which extends to infinity. This is the case, for example, in formula 2-167 where the exponential density disk is integrated from zero to infinity. However, the velocity curve obtained from the exponential distribution is not flat. If the density is assumed to drop to zero outside a given radius, the velocity curve then gets flatter. This is what is calculated above and also reported by Nicholson\cite{Nicholson2003b}. Note that Binney and Tremaine write on page 72: ``In this respect disks differ from spherical distributions of mass, for which the force at $r_0$ depends only on the density at $r < r_0$. In fact, the surface density of a disk at $R > r_0$ affects the attraction at $r_0$ because the annulus of material exterior to $r_0$ actually pulls a star placed at radius $r_0$ outward, thus partially compensating the inward attraction of the interior matter. At points in a disk where little matter is pulling outward, for example \textbf{on the perimeter of a sharp-edged disk, the circular speed can be much higher than at the edge of the spherical body with the same total mass and radius}.''\cite{Binney1987} [text in bold added here for emphasis] This example, a sharp-edged disk, is the solution given above.

The increase in density seen at the edge of the galaxy in Fig. \ref{fig:columndensity} occurs as a consequence of the abrupt termination of the density function, but only for the disk distribution, not for a spherical distribution. This happens because gravity from a ring attracts matter (located inside the ring) outwards. Near the edge of the galaxy, there is no more ring-distributed matter outside $R_{max}$. This may be counter-intuitive, but adding a little bit of mass at the edge of the disk actually makes the rotation curve flatter. A sharp mass decrease at the edge is compatible with observations without the need for an ``infinite mass''.\cite{Nicholson2000} Since large quantities of atomic hydrogen gas are present beyond the visible edge of galaxies, the edge of the density distribution may not coincide with the radius of the ``light emitting'' edge. Oscillations of the density also tend to develop if the smoothing is not used. A strong smoothing flattens this oscillation, but otherwise, it seems that the arms of the galaxy could be the result of these oscillations. These are ignored here, as they would constitute a research project of its own. A study of the stability of a disk is beyond the scope of this paper.

At the ``galactic edge'', the mass density drops quickly to its intergalactic level. A spherically symmetric intergalactic density does not affect any of the dynamics within the galaxy. Therefore, the calculations above are valid for any spherically distributed mass distribution outside the galaxy.

The numerical model assumes that the density of matter rapidly falls to zero at $R_{max}$. This radius may be past the limit of the light emitting matter in the galaxy, or the edge of the absorbing matter in the galaxy. The use of a zero density beyond $R_{max}$ is justified by the appearance of many spiral galaxies seen almost on edge. These galaxies show a dark band of light absorbing matter at the outer limit of the disk. However, careful examination shows that this light absorbing material is present up to a certain radius since no light absorption is present outside this radius. The disk of the Sombrero galaxy, for example, has a dark edge which absorbs the light emitted by other, more central, parts of the galaxy. Clearly, one can see that the density of the absorbing material stops abruptly at a certain radius $R_{abs}$: the light from the bulge is visible below the edge of the disk at the center of the picture, showing a transparent disk beyond $R_{abs}$.

{\bf Thermal Pressure}

The gas component of a galaxy is a mixture of atoms and molecules. These particles interact mostly with each other in elastic collisions.\cite{Governato1996} The solid matter component of a galaxy is a mixture of dust, rocks, planets and stars. These objects also collide with each other and with the gases, but because the collisions are inelastic, they contribute to essentially no pressure (equivalently, the temperature of the solid matter is very low). Since there is very little pressure, the density distribution of solid matter collapses to a two-dimensional disk. If the disk of the galaxy is supported by centrifugal forces, the bulge may seem more intriguing. At the poles where no rotation is measured (but speed distributions are observed), a thermal pressure is needed to provide the force to support the bulge from being pulled inwards by gravity. A velocity dispersion exists since the central bulge shows broadened lines, interpreted as orbits in random directions. However, if pressure is included in the model, a larger mass is expected so the gravitational force would become stronger to balance the outward pressure force. One concludes that the two components of a galaxy, solid and gaseous matter, will naturally separate in a disk and a bulge, respectively.  This correlates with the observations of heavier elements found in the disk (Population I system).

If any pressure effect is to be considered, thermal pressure is the largest for most galaxies. Radiation pressure would arise through a light scattering mechanism or photon absorption. Since light emitted from parts near the center of the galaxy makes it through to the observer, absorption and scattering events are rare, with a mean free path larger than several $kpc$. Otherwise, one would not see the galaxy's inside but just a blur.  However, collisions between atoms (called here thermal pressure) are more frequent. Even at a density of $1$ molecule$/m^3$, the mean free path between collisions is $3\ kpc$. A collision radically changes the momentum of these atoms, while a photon scattering event changes the momentum of an atom by a very small amount. Based on this argument, radiation pressure is negligible compared to thermal pressure.

\begin{figure} 
\centerline{\includegraphics[width=5.5in]{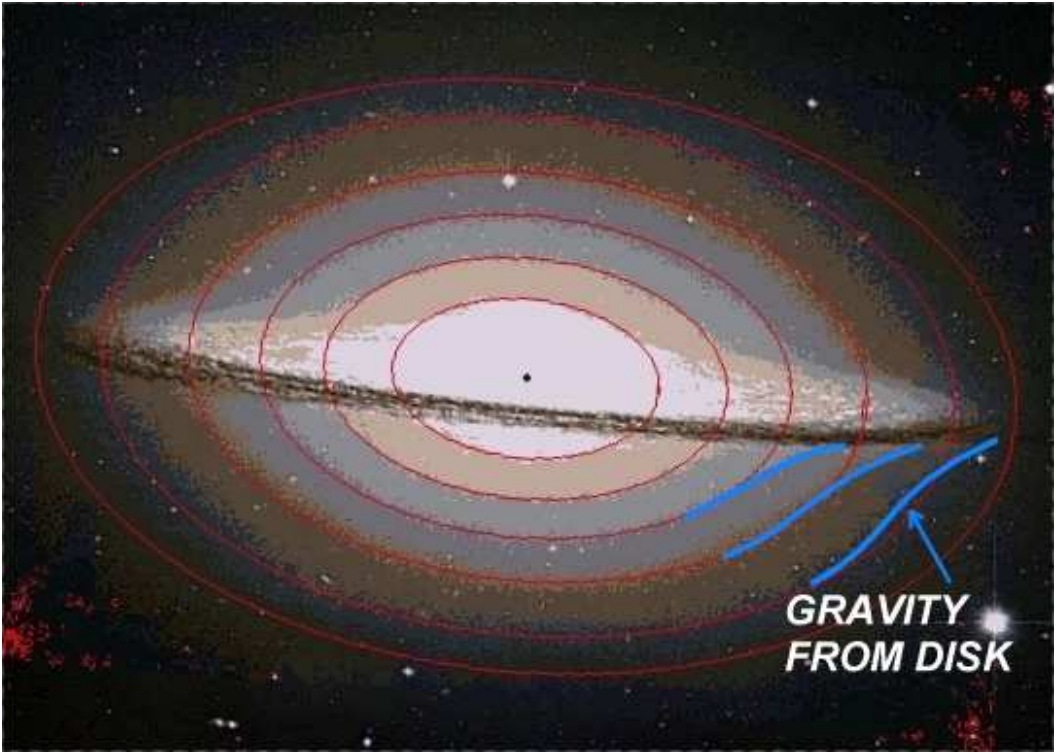}}
\caption{The Sombrero galaxy with elliptical isophotes. Near the galactic plane, the ellipses are modified by the gravitational field of the disk.}
\label{fig:sombreroisophotes}
\end{figure}

With an internal pressure supporting the gas in a galaxy, the general shape of the density function will be ellipsoidal. This is apparent on the picture of M104 in the visible represented using only 16 colors, as shown in Fig. \ref{fig:sombreroisophotes} - the ellipsoidal isophotes become very apparent. If no rotation is present, the density function is spherical (E0 galaxy).

{\bf Plasma effects}

It is known that plasmas contribute to the formation of galaxies in their early development\cite{Peratt1986a, Peratt1986b, Peratt1990}. The presence of plasmas in galaxies produces magnetic and electric fields which in turn affect the plasma at other locations in the galaxy. However, the plasma content of older galaxies is only a small fraction of the total mass of the galaxy. This is the justification used to neglect plasma effects in the current model.

%
%
%


%
{\bf Surface luminosity, Mass-to-Luminosity Ratio and Dark Matter}

The surface luminosity is derived from the column density Fig. \ref{fig:columndensity} assuming a constant mass-to-luminosity ratio and negligible absorption. For $1kpc < R < 4kpc$, the luminosity resembles the isothermal model $I(R) = I_s/(1+(R/R_0)^2)$ with $R_0 \approx 2kpc$. For $R > 4kpc$, the luminosity nearly follows the exponential law $I(R) = I_e exp(-R/R_d)$ with $R_d \approx 7kpc$. If the mass-to-luminosity ratio is $M_{sol}/L_{sol} = 0.05$ is used, a good fit is obtained with the typical luminosity measured in galaxies. The luminosity does not vary by more than a factor two for all the possible values of the angular momentum. This is consistent with the observation that the central surface brightness of a spiral galaxy is remarkably constant\cite{Binney1987} at $I_0 = 140 L_{sol}/pc^2$.

The density calculated here is in agreement with experimantal values\cite{Valentijn1999}. The small mass-to-luminosity ratio is the result of large quantities of "dark matter".  However, there are many possible candidates for dark matter that do not involve exotic non-baryonic matter.  Of these candidates, molecular hydrogen is the most likely to have enough mass to resolve the problem of the missing matter. This massive component, present in spiral galaxies, dominates the potential and the rotation curves\cite{Valentijn1999}. Other ``invisible'' condensed matter in the form of centimeter sized rocks to asteroid sized bodies are also possible. This matter would not have a distinctive optical signature and would have a very small optical cross section, since for any solid material of typical macroscopic radius $r$, the cross section increases as $r^2$ while the mass increases as $r^3$. Dark matter distributed on the disk and bulge is sufficient to explain the dynamics of spiral galaxies\cite{Mendez2001, delRio2001, Bertin1994}.

Binney and Tremaine\cite{Binney1987} dismiss $H$ and $H_2$ from work by Gunn and Peterson (1965) who looked at quasar light and absorption of that light in the interstellar gas. They conclude that very small amounts of $H$ or $H_2$ are possible, otherwise there would be much more absorption of the light coming from quasars. But the argument assumes that quasars are at the large distances they appear to be at. Some quasars are however believed to be closer\cite{Bell2006}. In this case, the real density of $H$ and $H_2$ can reach the values estimated byValentijn et al.\cite{Valentijn1999}.

\section{Conclusions} \label{sec:conclusions}
Although other models produce a flat rotation curve, the present paper also suggests an explanation for the low experimental mass-to-luminosity ratio. As opposed to the distributions with infinite extent used by Binney and Tremaine\cite{Binney1987}, a limited distribution is used in this paper. Given the rotation curve, the intrinsic angular momentum, the maximum radius and a smooth mass distributions as boundary conditions, a unique mass distribution is obtained. The mass distribution is finite and does not have to fit a simple analytical function. The additional mass needed to obtain an agreement with the observed mass-to-luminosity ratio is provided by baryonic matter most likely to be molecular hydrogen and condensed matter.

The argument put forward in a recent article\cite{Rubin2006} is frequently used in the scientific literature in order to support the hypothesis of dark matter. It is puzzling how this flawed argument has survived so long, when in fact the low mass-to-luminosity ratio in galaxies provides the main argument in favour of "matter that has no light". Vera Rubin's experimental work will certainly be useful to provide excellent data supporting the theoretical arguments of the current paper.

Much remains to be done. This paper does not address the problem of dark matter inferred from galaxy motion in clusters. Also, calculations with plasmas also produce matter distributions which explain galaxy dynamics. It would be important to know which fraction of interstellar matter is a plasma and contributes as such.

\bibliography{paper}
\bibliographystyle{spiebib}

\section{Appendix A: Numerical Integration} \label{sec:numericalintegration}

\begin{figure} 
\centerline{\includegraphics[width=5.0in]{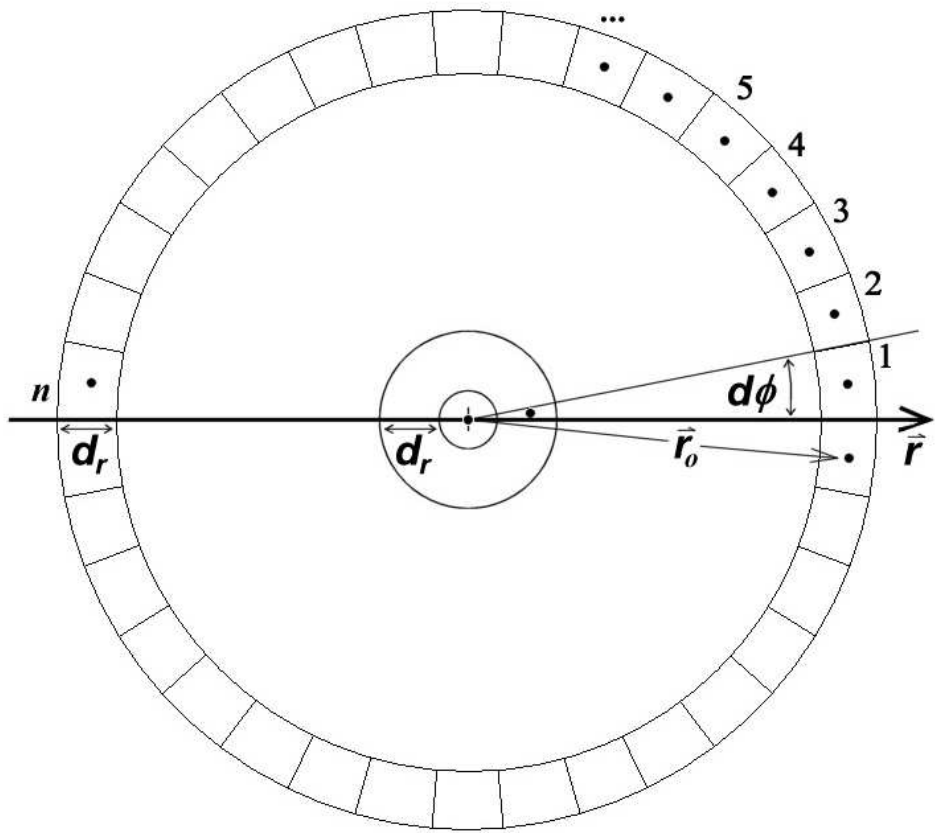}}
\caption{The numerical integration over the $\phi$ variable avoids the discontinuity problem by chosing sectors symmetrically about the $\vec r$ axis.}
\label{fig:phiintegration}
\end{figure}

It is important to properly handle the numerical integration of the rings to avoid discontinuities. The method used here is shown graphically in Fig. \ref{fig:phiintegration}. The sectors defined by the $\phi$ variable are offset from the $\vec r$ axis.  The mass is taken to be at the center of the sectors, as shown by the black dots.  This gives a non-null distance between all points $\vec r$ and $\vec r_0$.

The numerical integration is performed on $n$ elements on one side of the $\vec r$ axis, with $\phi = 0$ to $\pi$.  The acceleration along $\vec r$ is then obtained by doubling the result of the integration, while the acceleration perpendicular to the direction of $\vec r$ is always null.

The radial integration starts with an element at the center having a radius $d_r/2$.  The following elements all have a width $d_r$.

\end{document}